\documentclass[aip,prb,amsmath,amssymb,reprint]{revtex4-1}
\usepackage{graphicx}
\usepackage[version=3]{mhchem} 
\usepackage{subfig}
\usepackage{dcolumn}
\usepackage{bm}

\begin{document}

\title{Ferroelectric Materials for Solar Energy Conversion: Photoferroics Revisited}

\author{Keith T. Butler}
\author{Jarvist M. Frost}
\author{Aron Walsh}
\email[Electronic mail:]{a.walsh@bath.ac.uk}
\affiliation{Centre for Sustainable Chemical Technologies and Department of Chemistry, University of Bath, Claverton Down, Bath BA2 7AY, UK}

\date{\today}

\begin{abstract}
The application of ferroelectric materials (i.e. solids that exhibit spontaneous electric polarisation) in solar cells has a long and controversial history. 
This includes the first observations of the anomalous photovoltaic effect (APE) and the
bulk photovoltaic effect (BPE).
The recent successful application of inorganic and hybrid perovskite structured materials (e.g. \ce{BiFeO3}, \ce{CsSnI3}, \ce{CH3NH3PbI3}) in solar cells emphasises that polar semiconductors can be used in conventional photovoltaic architectures. 
We review developments in this field, with a particular emphasis on the materials known to display the APE/BPE (e.g. ZnS, CdTe, SbSI), and the theoretical explanation. 
Critical analysis is complemented with first-principles calculation of the underlying electronic structure.
In addition to discussing the implications of a ferroelectric absorber layer, and the solid state theory of polarisation (Berry phase analysis),  design principles and opportunities for high-efficiency ferroelectric photovoltaics are presented.  
\end{abstract}

\maketitle 

\section{Introduction}
Ferroelectrics are a class of materials that display spontaneous electric polarisation. This is due to the breaking of centrosymmetry of the crystallographic unit cell, and may be varied by the application of physical, chemical or mechanical bias. Ferroelectric materials have extensive potential technological applications, due to the possibility of coupling the ferroelectric response with other properties.  
Applications include memory storage media\cite{hu-2009,Scott_science_2007}, field effect transistors and ferroelectric random-access memory\cite{garcia-2012n,Lee-2012-adm}.  
The coupling of ferroelectricity and magnetism has led to an extremely fertile area of research, `multi-ferroics', with potential use in emergent spintronic technologies \cite{garcia-2010,ramesh-2010,meyerheim-2011}. 
An important realised technological function of ferroelectrics is the coupling of mechanical response to electric field, applied in both sensors and actuators.

Light-to-electricity energy conversion in ferroelectrics was envisioned 35 years ago by V. M. Fridkin, who imagined a ``photoferroelectric crystal'' as a potential solar cell\cite{fridkin-1979}. 
In the following decades the development of ferroelectric based photovoltaic (PV) devices has mostly remained the preserve of academic research.
Industry adoption is hampered by low quantum efficiencies of devices, as well as poor bulk conductivity of common ferroelectric materials. 
Further,  the theoretical description of polar properties in bulk materials remained incomplete until formalisation in the modern theory of polarisation\cite{king-smith-1993}. 
There have been significant recent advances in ferroelectric photovoltaics\cite{Seidel-2014r}, most notably in devices based on oxide and halide perovskites. 

In this perspective we consider a class of systems where the ferroelectric effect and photo-response are intimately linked: photoferroics. 
We will outline the ferroelectric and photovoltaic action, followed with an examination of the application of ferroelelectrics to solar cells, discuss several proposed models for enhanced PV performance observed in ferroelectric materials, and consider contemporary research into photoferroics. 
We will investigate a historically important but latterly overlooked class of photoferroic materials, the antimony chalcohalides. 
The perspective concludes with a consideration of new directions for materials design, and how ferroelectric materials can be applied in novel device architectures to improve photovoltaic performance. 

\section{The Ferroelectric Effect}

Pyroelectric crystals possess a net dipole moment (\textbf{P}) in their primitive unit cell, and therefore exhibit spontaneous polarisation. 
They generate a transient voltage when heated (hence `pyro') due to changes in lattice polarisation arrising from thermal expansion. 
All pyroelectric materials must adopt a non-centrosymmetric crystal structure, with an asymmetric (negative) electron density around the (positive) nuclei. 
These asymmetric charge density crystals are necessarily described by polar crystallographic point groups (10 out of the 32 groups), where more than one site is unmoved by every symmetry operation. 
This excludes all cubic crystals, and other high symmetry space groups. 
The electric dipole resulting from the ionic positions is defined as:
\begin{equation}
\textbf{P}_{ionic} = \sum_i q_i u_i
\label{p1}
\end{equation}
where $q_i$ and $u_i$ represent the ion charge and position, and the sum is over all atoms in the unit cell. 

Ferroelectrics can be defined as the subgroup of pyroelectric materials in which the equilibrium structure has no net dipole above a certain temperature. 
The orientation of spontaneous dipoles below this transition temperature can be manipulated by the application of an electric field---ferroelectric materials exhibit a polarisation which is dependent on the history of applied field (hysterisis). 
The first report of ferroelectricity was from Valasek, who recorded the response of Rochelle salt (\ce{KNaC4H4O6} $\cdot 4$ \ce{H2O})\cite{valasek-1921}.

In many cases the polar structure of a ferroelectric material may be obtained from non-polar polymorphs by a small mechanical strain, temperature change, or even a variation in carrier concentration. 
The free energies of the polar and non-polar polymorphs of a crystal are generally quite close (several meV per atom). 
A critical temperature exists (the Curie temperature, $T_C$) at which the free energies of both phases are degenerate. 
The polar (lower symmetry) structure is usually the low temperature ground state. 
 
When a material is in its pyroelectric phase it commonly consists of domains -- regions of homogeneous polarisation -- that differ only in the direction of the polarisation. 
For greater detail on the ferroelectric (and related piezoelectric) effect, several excellent text books are available\cite{kanzig,lines-1977,fridkin-1979}. 

\subsection{Modern theory of polarisation}
The interpretation of crystal polarisation was fundamentally altered 25 years ago by the modern theory of polarisation\cite{king-smith-1993,souza-2002,umari-2002}. 
The classical polarisation resulting from the position of charged ions in a lattice is well defined (Equation \ref{p1}), while the calculation of electric polarisation from periodic electronic wavefunctions posed a major theoretical challenge. 
There is no unique way to separate the charge density into finite regions of well defined polarisation.
By recasting the problem from real to reciprocal space, and applying Berry phase analysis\cite{berry-1984} to the change in phase of the electronic wavefunction summed over all wavevectors, an intrinsic polarisation can be directly and unambiguously computed for a periodic material. 

Berry's Geometric phase analysis has been applied in numerous studies revealing hitherto unknown aspects of ferroelectric materials. 
For electronic structure calculations such analysis requires a relatively low-cost post-processing of pre-computed electronic structure and ion positions. 
The total change in polarisation for a (ferroelectric) transition can now be defined as a sum of the ionic and electronic components:
\begin{equation}
\Delta \textbf{P} = \Delta \textbf{P}_{ionic} + \Delta \textbf{P}_{electronic}.
\label{p2}
\end{equation}

Recent applications include unusual ferroelectric instabilities in fluoroperovskites \cite{garcia-castro-2014}, highlighting the importance of covalent bonding in the piezoelectric response of \ce{BaTiO3}\cite{shi-2014-prb}, as well as the discovery of new classes of proper and improper ferroelectrics \cite{garrity-2013,young2014improper}. 
The approach has also been applied to the study of less traditional solid-state materials such as metal-organic frameworks\cite{stroppa-2013-adm,stroppa-2011}.
Building upon the work of von Baltz and Kraut\cite{baltz-1981a}, Rappe and co-workers were able to calculate the so-called shift-current contribution to photovoltaic performance\cite{young-2012b,young-2012a}. 
In the field of hybrid halide perovskites, Berry phase analysis has been used to demonstrate the possibility of molecular tuning of the electric polarisation\cite{frost-2014-nano}.

\section{The Photovoltaic Effect}
In a semiconducting material, the absorption of photons with energies above the band gap ($h \nu \geq E_g$)  results in the promotion of electrons from the valence band to the conduction band.
The process generates hole (valence band, effective positive charge) and electron (conduction band, negative charge) carriers. 
In a typical material these excited carriers will decay back to the ground-state, energy being conserved by the emission of light (photons, radiactive decay) or heat (phonons, non-radiative decay). 
 The photovoltaic effect occurs where an asymmetry in the electric potential across the material (or selective electrical contacts) results in a net flow of photogenerated electrons and holes: a photocurrent. 
 In one of the earliest examples of a solar cell, an asymmetric potential was introduced by placing a layer of selenium between two different metallic contacts\cite{nelson-2003}. 
 The difference in workfunctions of the metals creates the necessary asymmetry and electrically rectifying action, a Schottky barrier.
 
 In the 1950s an alternative method of creating asymmetry for charge separation was discovered. 
 By doping different regions of a single piece of silicon with phosphorous and boron, it is possible to establish an asymmetric potential in a single crystal, the $p-n$ junction. 
 Here the built-in field gives better rectification, and therefore better photovoltaic action. 
 The ideal photovoltaic material should separate charges as efficiently as possible, with minimal relaxation of the charge carriers from the optical excitation, and transport them independently to the contacts, thus minimising recombination (the loss pathway) between electrons and holes. 
 
 The overall power conversion efficiency ($\eta$) of incident light power ($P_{in}$) to electricity is is proportional to the open-circuit voltage ($V_{oc}$), short-circuit current ($J_{sc}$) and the fill-factor ($FF$):
 \begin{equation}
 \eta = \frac{V_{oc} J_{sc} FF }{P_{in}},
 \end{equation} 

 $V_{oc}$  is the potential difference developed across a cell (in the light) when the terminals are not connected (no current flow). 
This represents the maximum voltage which can be generated by the cell.
$V_{oc}$ is limited (in a standard semiconductor junction) by the band gap of the absorber layer, with additional recombination losses (at open circuit, all photo-generated charges recombine). 
$J_{sc}$ is the photo-current extracted when the voltage across the cell is zero, all generated potential difference is used to extract charge carriers. 
$J_{sc}$ is limited by the proportion of the solar spectrum absorbed by the active material in a device. 
At both $J_{sc}$ and $V_{oc}$, no energy is extracted from the solar cell. 
In an idealised device the power generation would equal the product $V_{oc} \times J_{sc}$. 
Detailed balance requires that radiative recombination must occur. 
Any additional recombination is a loss pathway. 
The realised power at the maximum power point on the $J-V$ curve as a fraction of the idealised power is the fill-factor, $FF$.

Recombination of photo-generated electrons and holes limits the efficiency of operating solar cells. 
Recombination can occur directly from valence to conduction band, or via trap states. 
Trap mediated recombination occurs when imperfections in the crystal cause a localised density of states within the band gap. 
Both electrons and holes can become energetically trapped, then recombining with carriers of the opposite sign. 
Surfaces represent a major source of trap states in convential semiconductors, with under-coordinated atoms introducing localised states into the gap. 
Surface effects can be reduced by the inclusion of a passivation layer, to satisfy coordination at the surface, whilst not conducting charge themselves\cite{butler-2011,lamers-2012}. 
Band to band recombination occurs when carriers of opposite sign encounter each other in the semiconductor. 
This is reduced by improving the mobility (reducing time in the device) separation (segregating oppositely charged carriers with an electric field) or screening of the carriers (reduces recombination cross-section). 

There are several photovoltaic architectures which are used to achieve efficient charge separation and transportation. 
$p-n$ homojunctions as outlined above; $p-n$ heterojunctions (e.g. in CdTe cells), which are similar to standard $p-n$ junctions, but consist of two distinct materials; $p-i-n$ junctions (e.g. in $a$-Si cells) have a region of undoped (intrinsic) material between the $p-$ and $n-$ regions. 
Organic solar cells typically require a heterojunction to efficiently drive the charge separation of tightly bound excitons, sacrificing photon generated. 
This consists of electron donor and acceptor molecules in close proximity, with hole- and electron- selective electrodes.

\begin{figure*}[ht!]
\begin{center}
\resizebox{15 cm}{!}{\includegraphics*{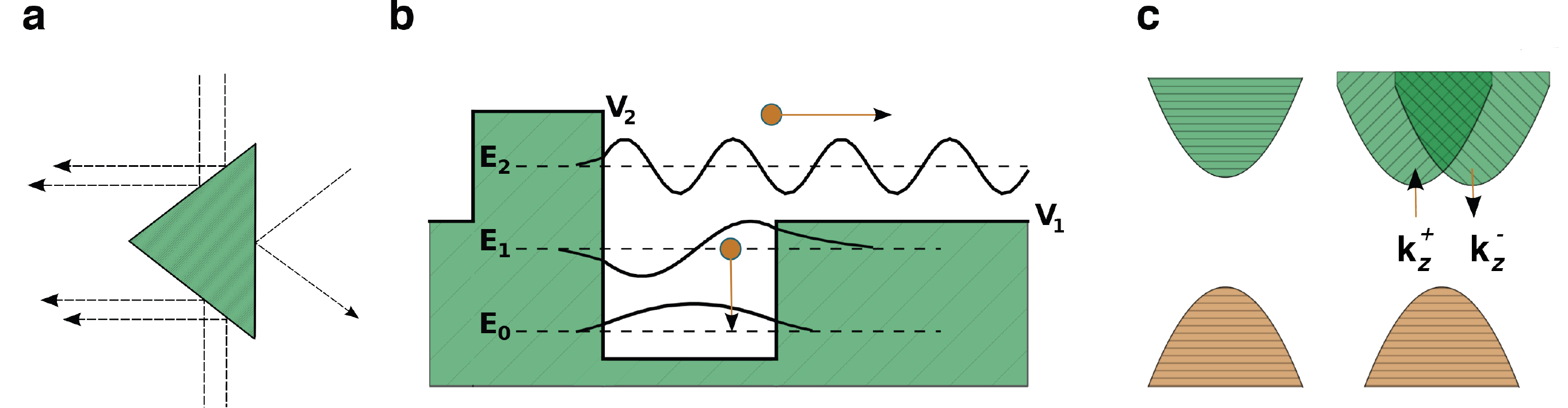}}
\caption{\label{Bulk-photo} Models for the bulk photovoltaic effect. a) Asymmetric carrier scattering centres, resulting in a net flow from randomly drifting carriers, following Belinicher\cite{belinicher-1980}. b) Asymmetric potential well at a carrier generation centre;  photogenerated carriers have a preferred direction of exit, following Lines\cite{lines-1977}. c) Relativistic splitting of the conduction band minimum establishes two distinct channels for electron excitation, polarised light promotes electrons preferentially to one channel, following Fridkin\cite{fridkin-2012}.}
\end{center}
\end{figure*}

\section{Ferroelectric Photovoltaics}
We have outlined common photovoltaic device architectures, which universally rely on charge separation by variation in material composition. 
Charge separation due to the innate crystal field in a homogeneous material is also possible, which is the process used in some ferroelectric photovoltaics.  
The crystal polarity creates microscopic electric fields across domains, separating photogenerated excitons into free charges, and segregating the transport of the free charges to reduce recombination rates.


There are additional potential advantages to such device designs. 
For example, ferroelectric materials can achieve extremely high open circuit voltages (V$_{oc}$), unlike a standard photovoltaic cell where V$_{oc}$ is limited by the band gap of the absorber material. 
The consistency of the product of $J_{sc}$ and $V_{oc}$ is maintained with larger photovoltages being associated with smaller photocurrents.


Recent research into ferroelectric photovoltaic materials has consisted of two mostly independent strands.
Photovoltaic effects have been studied in oxide ferroelectrics\cite{yang-143,grindberg-509,alexe-2011,bhatnagar-2013,young-2012b}, notably BiFeO$_3$ (BFO), from a fundamental physics and materials design perspective.
The discovery that hybrid organic-inorganic halide perovskites, notably CH$_3$NH$_3$PbI$_3$ (MAPI), can make high-efficiency photovoltaic devices has redirected vast areas of solar energy research\cite{kojima-6050,snaith-1,gratzel-1,heo-486,park-2423,bass2014influence,bhachu2015scalable}. 
The former has been driven by both the layer-by-layer control of modern deposition techniques and the development of the modern theory of polarisation facilitating a complete description of bulk polar materials. 
The latter has been driven by the extraordinary empirical performance of MAPI, demonstrating the potential of polar materials.

\subsection{Photoferroic Phenomenology}

The photoferroic current ($J_i$) can be related to absorbed light by the rank three tensor $\beta$:
\begin{equation}
J_i = p_j p_l^* \beta_{ijk} I_o
\end{equation}
where $I_o$ is the intensity of the absorbed light (assuming isotropic absorption) and $p_n$ is the polarisation of the medium, the subscripts $ijk$ correspond to the direction in space. The amplitude of the tensor has the form

where $I_o$ is the intensity of the absorbed light (assuming isotropic absorption) and $p_n$ is the polarisation of the medium in direction $n$. 
The amplitude of the tensor has the form

\begin{equation}
\beta_{ijk} = e l_o \zeta \phi (\hbar \omega)^{-1}
\end{equation}
where $\zeta$ describes the excitation asymmetry, $\phi$ is the quantum yield, $\hbar \omega$ is the photon
energy. 
and $l_o$ is the mean free path of excited carriers and $e$ is the elementary charge. It can be shown\cite{zenkevich-2014} that the
 efficiency of power conversion from the photoferroic effect can be expressed as :
 \begin{equation}
 \eta = \beta _{ijk}E
 \end{equation}
where E is the electric field arising from $j_i$
\begin{equation}
E = \frac{J_i}{\sigma_{pv}}
\end{equation}
where $\sigma_{pv}$ is the photoconductivity.

Generally, in bulk crystals the values of $\beta$ and $E$ are very small. 
When the size of the sample is of the order of $l_o$, all of the excited carriers contribute to photo-current and $E$ can become much larger. 
Within band theory, the length has been estimated to be 10--100 nm\cite{fridkin-1992}; hence, photoferroic effects are enhanced at the nanoscale.
Understanding and controlling $l_o$ is an important aspect in the design of photoferroic device architectures.

In a photoferroic system there is an intricate relationship between photo-response and ferroelectric phase stability (including domain size and distribution). 
The relatively high concentration of photo-generated carriers (electrons and holes) means that the electronic subsystem has an appreciable effect on free energy close to ferroelectric transition (Curie point). 
The electron subsystem can alter the nature of these phase changes, which can be used to experimentally classify a system as a photoferroic as well as to quantify the photoferroic effect.

In general, \textit{temperature hysteresis} is reduced and \textit{Curie-points} are lowered in the presence of photo-excited charges; the shifts are proportional to carrier concentration, as observed for example in SbSI\cite{belyaev-1967}. 
 \textit{Spontaneous polarisation} as measured, for example by pulsed-field or hysteresis loop methods, is reduced by the presence of free charges due to enhanced screening. 
\textit{Structural deformation} can be caused by the presence of carriers, with charges affecting the unit cell volume during the phase transition. 
\textit{Effective permittivity} has a dependence on the presence and concentration of carriers; the dielectric screening is initially increased by increasing carrier concentration. 
This outline of the physical manifestations of the photoferroic effect is necessarily limited, for a comprehensive review of these properties, as well as the thermodynamic principles underlying them we refer the reader to V. M. Fridkin's seminal texts\cite{fridkin-1979,fridkin-1992}. 

Numerous mechanisms have been proposed to explain the unusual photovoltaic performance of ferroelectric materials. 
We now describe several of the key models used to rationalise experimental phenomena in poly- and mono-crystalline materials.  

\subsection{Bulk Photovoltaic Effect (BPE)}

Photovoltages in un-doped, single crystal samples of materials have been reported as a bulk photovoltaic effect (BPE), sometimes referred to as the photogalvanic effect or non-linear photonics. 
The earliest report was of steady-state photovoltages in single crystal BaTiO$_3$ (BTO) in 1956\cite{chynoweth-1956} and it is only observed in non-centrosymmetric systems. 
The recorded photocurrents were closely related to the magnitude and the sign of the macroscopic polarisation of the sample. 
Subsequently, similar effects were reported in LiNbO$_3$ and LiTaO$_3$\cite{chen-1969}. 
More recently there have been a series of studies on BiFeO$_3$ (BFO), with intense interest in its application as a photoferroic material in PV devices. 

\begin{figure*}[ht!]
\begin{center}
\resizebox{15 cm}{!}{\includegraphics*{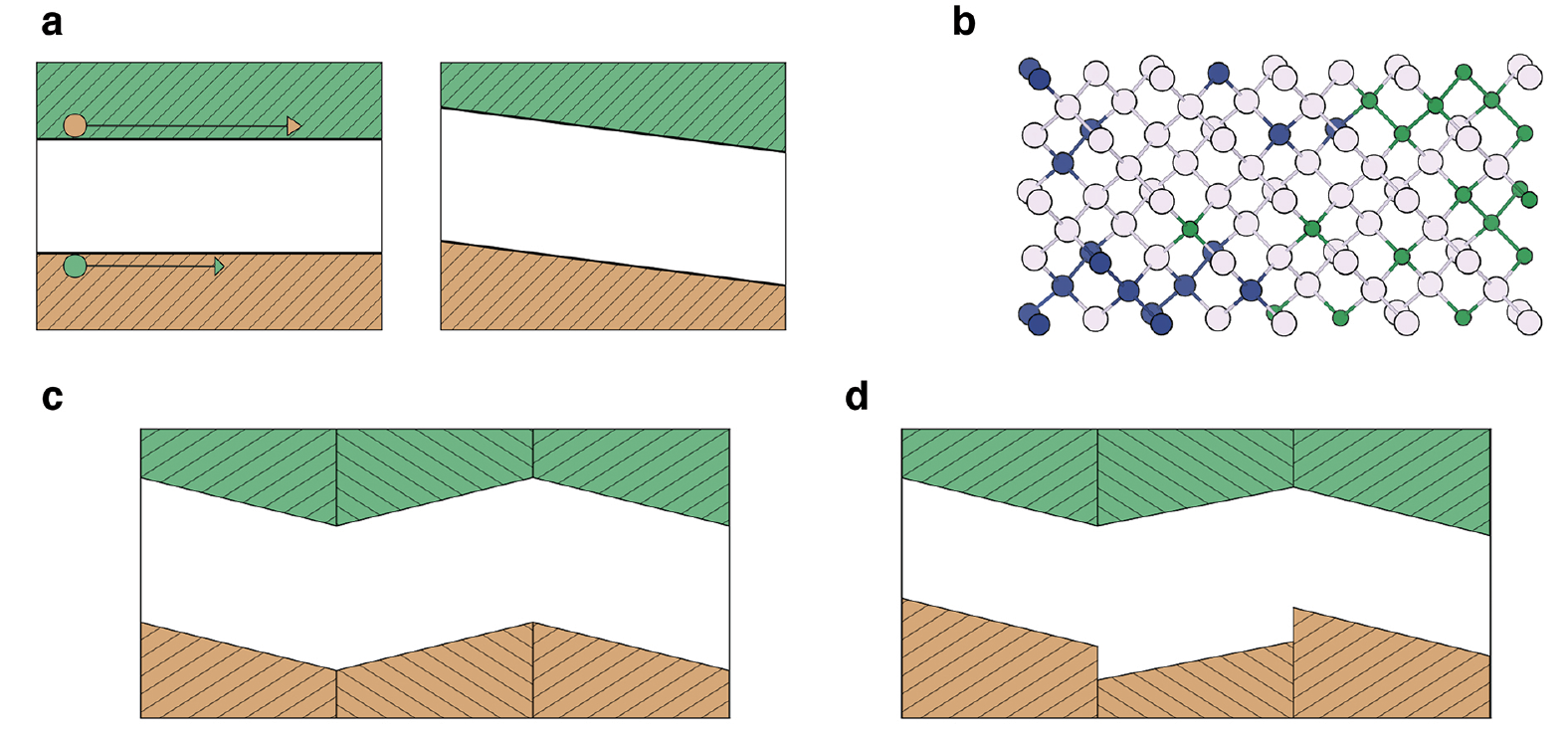}}
\caption{\label{Poly-photo} Models for the poly-crystalline anomalous photovoltaic effect. 
Valence and conduction bands are coloured orange and green, respectively. 
(a) The Dember effect: holes and electron have different mobilities, resulting in a net charge 
(with an internal electric field) across the crystal upon carrier generation. 
(b) Asymmetric aliovalent doping of a crystal results in an electrostatic potential across a grain. 
(c) Breaking of centro-symmetry creates a dipole in the crystal unit cell and can result in ferroelectric domains.
(d) If a poly-crystal contains inhomogeneous domains, the gradients in electric potential do not exactly cancel and a voltage is generated across the poly-crystal that may exceed the band gap of the material.} 
\end{center}
\end{figure*}

The simplest proposed model for the BPE is based on asymmetric scattering centres in a material\cite{belinicher-1980}, which is graphically represented in Figure ~\ref{Bulk-photo}a. If a medium contains randomly located, but identically oriented wedges, then, in the absence of external forces, the random diffusion and drift of carriers in the medium will eventually set up a net current. Any current established by this mechanism, however,  would be expected to be local and short lived, constrained by the associated increase in entropy.  


Another model, based on asymmetry in the electrostatic potential in which electrons and holes diffuse is detailed in Figure ~\ref{Bulk-photo}b \cite{lines-1977}. 
In this case there is anisotropy in the potential at an absorbing centre, for example, caused by crystal polarisation.
 Carriers are excited from the state at $E_0$ to an energy $E$.
 If  $E < V_1$ the excited electron remains trapped in the potential well.
 If  $E>>V_2$ then the carriers move away from the centre isotropically; however, if $V_1 < E < V_2$ then carriers moving to the left are partially scattered by the potential barrier (allowing for a certain probability of tunneling) and a net flow of carriers to the right (as indicated in Figure ~\ref{Bulk-photo}) is established. 
Thus optical absorption in a polar crystal with aligned asymmetries results in a net current. The contribution of this current to the overall photocurrent is maximized when the width of the crystal is similar to the decay length of the asymmetrically excited carriers\cite{zenkevich-2014}.
Due to a decay length of 10 -- 100 nm\cite{zenkevich-2014}, as discussed previously, the effect is maximised in ultra-thin films.

A third model relates to the BPE in gyrotropic crystals\cite{fridkin-2012}, materials whose valence or conduction bands are split in reciprocal space by relativistic spin-orbit coupling (e.g. so-called Dresselhaus or Rashba splitting\cite{dresselhaus1955spin}), Figure ~\ref{Bulk-photo}c. 
A key requirement again is the absence of crystal inversion symmetry. 
In a non-relativistic description, all valence band electrons have an equal probability of being excited by photons. 
When the bands are split by spin-orbit coupling, the momentum of the excited carrier depends on its spin. 
Therefore illumination by polarised light results in current flow. 
Clockwise polarised light excites electrons to a state momentum $k_z > 0$ and anti-clockwise polarised light excites electrons to $k_z < 0$, because of the decoupling of the spin channels. 
The nature of the splitting may be different for electrons and holes due to the different orbital contributions to the conduction and valence bands.
These spin-orbit coupling effects are larger with heavier elements (e.g. Pb and Bi), and has been recently demonstrated for the hybrid perovskite MAPI\cite{even-2999,brivio-2014}.

The BPE mechanisms considered above arise from the asymmetric velocities of carriers in the potential of the crystal lattice. Another important contribution is due to the asymmetry of the electron density. This results in excitation of carriers in one band to another band, which is separated from the initial one in real space. These ``shift currents'' have been demonstrated for several materials including \ce{BiFeO3}\cite{young-2012a},\ce{BaTiO3}\cite{wei-2010} and \ce{GaAs}\cite{sipe-2000,bieler-2007}, both experimentally and theoretically. Theoretical results for \ce{BaTiO3} demonstrate that for a significant shift current a material must feature `covalent' bonding that is highly asymmetric along the current direction\cite{young-2012b}.  

\subsection{Anomalous Photovoltaic Effect (APE)}

The photovoltage achievable from a semiconductor is generally limited by the bandgap of the light absorbing material. Starkiewiz and co-workers first reported observations contravening this general rule on PbS films in 1946\cite{starkiewicz-1946}. Subsequently similar observations were reported for polycrystalline CdTe, ZnTe and InP\cite{johnson-1975,goldstein-1959,uspenskii-1968}. 
The common feature was thin films deposited on an angularly inclined and heated substrate. 
Reports of photovoltages hundreds and even thousands of times the bandgap were highly sensitive to the conditions of the samples and were extremely difficult to reproduce. 
A coherent model explaining the effect was slow to emerge. 
An unusual aspect is that the materials mentioned above are not known to be ferroelectric, e.g. PbS adopts the rocksalt structure, which is stable up to high temperatures\cite{skelton}.

The explanations which were put forward generally fall into three categories:  
(a) the Dember effect;
(b) $p-n$ homojunction domains; 
(c) ferroelectric domains. 
All three explanations posit an inhomogeneity in the charge distribution, which is not fully screened due to the presence of crystal nano- or micro-structure. 
The resulting photovoltage across the material can be additive depending on the number and type of domains/interfaces present. 
The mechanisms, outlined in Figure~\ref{Poly-photo}, can be summarised as follows: 

(a) In the Dember effect (Figure ~\ref{Poly-photo}a)  photo-generated charge carriers are generated inhomogeneously throughout the crystal, forming preferentially at the face exposed to the photon source\cite{dember-1931}. 
Carriers then diffuse through the material; however, the difference in effective masses of the electron and hole carriers means that diffusion occurs at different rates, thus a net polarisation across the material is developed. 

(b) In the $p-n$ junction model (Figure ~\ref{Poly-photo}b) each crystallite is presumed to have an inhomogeneous distribution of $p$ and $n$ type defects or dopants, creating the difference in electric potential necessary to separate charge carriers in the crystallite\cite{starkiewicz-1946}. 

(c) In the ferroelectric domain picture (Figure ~\ref{Poly-photo}c), the charge carrier separation results from the polarisation of the material itself, in the form of the electric fields due to a ferroelectric domain structure\cite{ellis-1958}. 

In the above cases the steady-state photovoltage will not exceed the bandgap of the material. 
This is because in a single crystal the depolarisation field would be exactly cancelled by the formation of space-charge regions at the boundaries. 
In a poly-crystal there can be an array of alternating interfaces, e.g. AB BA (Figure ~\ref{Poly-photo}c). 
If either A or B is pyroelectric then AB and BA junctions are not equivalent.
In this case the space-charge may not fully counteract the depolarisation field and the resulting photovoltage can then build up across the poly-crystal resulting in an above bandgap potential difference (Figure ~\ref{Poly-photo}d)\cite{lines-1977}. 
Unlike the BPE, which can be defined as an intrinsic bulk response, the APE relies on the nano- and micro-structure of a sample.  
  
\section{Oxide Perovskites}
The study of ferroelectrics has been dominated by perovskite-structured metal oxides. 
These are ternary materials of the form \ce{ABX3}, where the A site in the crystal lattice is at the centre of a three-dimensional network formed by corner-sharing \ce{BX6} octahedra.
At high temperature, a high symmetry cubic structure is commonly observed (where polarisation is forbidden by inversion symmetry), while at lower temperature a range of lower symmetry phases can be formed (e.g. tetragonal, rhombohedral and orthorhombic perovskites), which can be ferroelectric or antiferroelectric.
The phase diagrams of these materials are highly complex, with a combination of short and long-range order (see for example recent work on \ce{PbZr_{1-x}Ti_xO3})
\cite{zhang2014missing}.

Most oxide perovskites are wide bandgap insulators, and high-temperature conductivity is usually ionic, rather than electronic\cite{catlow-3379,walsh2011strontium}.
For ionic-conducting perovskites, aliovalent doping can be performed to increase vacancy concentrations (to enhance mass transport) rather than electron or hole concentrations. 
Indeed even for hybrid halide perovskites, a strong preference for ionic compensation of charged point defects has been predicted\cite{walsh2014self}.

The few demonstrated solar cells based on oxide perovskites have poor power conversion. 
For example it has been shown for single crystal \ce{BaTiO3} that the conversion efficiency due to the BPE is limited to $\sim 10^{-7}$\cite{zenkevich-2014}.
Nonetheless there have been recent reports of improved performance by decreasing layer thickness and judicious engineering of domain and electrode interfaces.
While initial power conversion efficiencies were in the region of 0.5\%\cite{zenkevich-2014,grindberg-509},
there has been recent success up to 8\% for \ce{Bi2FeCrO6}\cite{nechache2014bandgap}.

One of the major hindrances faced by the oxide materials are the wide bandgaps, which allow for only a small fraction of the solar spectrum to be absorbed. 
There have been recent reports of bandgap-engineered materials with ferroelectric properties and bandgaps appropriate for efficient solar energy conversion\cite{wang-235105,wang-152903}.  
By controlling the cation ratio and distribution in the double perovskite \ce{Bi2FeCrO6} (Fe and Cr ions are distributed over the perovskite B site), the optical band gap could be tuned by several eV\cite{nechache2014bandgap}.

The authors who first reported above bandgap photovoltages in \ce{BiFeO3} dismissed the possibility that the phenomenon has the same origin as the bulk photovoltaic effect in other single crystals such as  \ce{BaTiO3}  and LiNbO$_3$\cite{yang-143}. 
In this instance the role of domain walls (the interface between ferroelectric domains) was emphasised by a series of experiments demonstrating the dependence of the obtained photovoltage on the domain wall density. They proposed a model building upon theoretical studies that demonstrated the presence of built-in electric potential at domain walls\cite{meyer-2002,seidel-2009}. 
Excitons (electron-hole pairs), which would otherwise be tightly bound in BFO, are separated by the electric field at the domain walls. 
In this case the charge generated at the domain walls acts to depolarise the field in that region, whist the field in the bulk material remains.
The imbalance of polarisation is responsible for the above bandgap photovoltages.

The first model developed for BFO has subsequently been disputed by other groups, who in a series of equally elegant experiments highlighted the independence of photovoltages on the domain wall density \cite{bhatnagar-2013}. 
The problem of understanding is due to the difficulty in separating bulk photovoltaic and polarisation-dependent mechanisms. Recent advances in the theory of polarisation mean that deciphering these contributions and interpreting experiments with the aid of first principles calculations has become possible.   
Rappe and co-workers calculated the shift-current tensor for BFO\cite{young-2012a}, revealing the anisotropy of the photo-induced currents. 
Due to distinct experimental set-ups, the measurements in references\cite{yang-143} and \cite{bhatnagar-2013} probe different orientations, which reconciles the initial disagreement between the studies. 

\section{Hybrid Halide Perovskites}
Hybrid halide perovskites have had a radical impact on solar energy research in the past two years, motivated by the highest power conversion efficiencies demonstrated for a low-temperature solution-processed semiconductor.  
Since their first reports as PV materials\cite{kojima-6050}, devices based on these materials have made enormous progress in mesoporous and thin-film configurations\cite{snaith-1,gratzel-1,bisquert-4,heo-486, snaith-2, snaith-7467,zhou-2014}.
The recent highest-efficiency device exceeds 20\% light-to-electricity conversion.


Halide perovskites have the same structure as the oxide counterparts, with the oxygen anions replaced by a halide. 
This change in the formal oxidation state of the anion means that to keep charge neutrality the oxidation state of the cations must sum to III (usually by the combination of monovalent and divalent species). 

For \ce{CH3NH3PbI3}, the upper valence bands are composed of I 5p states\cite{mosconi-13902,brivio-2014}, which results in a higher energy valence band (lower ionisation potential) than in the oxide perovskites. 
The B cation has thus far generally been Pb (with some reports of limited success with Sn\cite{noel-2014}), which results in large spin-orbit coupling, lowering the conduction band by a significant degree\cite{brivio-2014,umari-2014}. 
The combined effects of the higher valence band and lower conduction band means that the optical bandgap of the halide perovskites are significantly smaller than the oxide analogues\cite{brivio-042111}, allowing for efficient absorption of white light.
At the same time, the rich chemistry and physics associated with the perovskite structure is maintained. 

The distinction for \textit{hybrid} perovskites is that the \textit{A} site cation is an organic molecule as opposed to an inorganic ion. This introduces a number of important extra degrees of freedom. 
The crystal symmetry is directly reduced; even a cubic arrangement of the \ce{BX6} octahedra can  result in a net polarisation. 
The large static electric dipole of the methyl-ammonium (\ce{CH3NH3} or MA) ion, used in MAPI is suggested as on contributing factor to crystal polarisation and PV performance\cite{frost-2014-nano}. 
The orientational dynamics of the MA have also been implicated as effecting structural changes in the material\cite{gottesman-2014}. 
By varying the size of the organic molecule the bandgap can be manipulated\cite{brivio-042111}. 
Larger cations cause the break-up of the structure into 2D layers\cite{mitzi-1,calabrese-2328,borriello-235214}, the 3D perovskite structure is stable only with a small subset of possible ion choices.

One unusual aspect of the device physics of halide perovskite solar cells is significant hysteresis in the photovoltaic (J--V) response\cite{snaith-hysteresis-2014}.  
Two likely contributing factors are ion diffusion and ferroelectricity. 
Indeed the first direct observation of ferroelectric domains in \ce{MAPbI3} have just been reported\cite{kutes2014direct}, and first-principles calculations predict spontaneous electric polarisation similar in magnitude to inorganic perovskites\cite{frost-2014-nano}. 
A complicating factor is the orientational disorder of the dipolar MA ion, which is sensitive to temperature and lattice strain. 
Monte Carlo simulations for \ce{CH3NH3PbI3} have shown that a low temperature antiferroelectric structure of twinned dipoles becomes disordered at room temperature due to entropy; however, significant short-range order is maintained. 
This behaviour is consistent with the low-temperature orthorhombic ordered phase and the disordered tetragonal and pseudo-cubic 
structures observed around room temperature.
The shift-current for MAPI, computed from first-principles, suggests a BPE in the visible range approximately three times larger than oxide perovskites, and the effect is sensitive to molecular orientation\cite{zheng2014first}.

The channels of high and low electrostatic potential associated with the correlation between MA ions may 
provide efficient diffusion pathways for electrons and holes.  
The domain behaviour is highly sensitive to the presence of an external electric field.
The changes in the 3D electrostatic potential landscape could be linked to differences in electron-hole recombination rates under short-circuit and open-circuit conditions, and thus give rise to the observed hysteretic effect\cite{frost-2014-apl}.

\section{Beyond Perovskites: Chalcohalides}
\begin{figure}
\begin{center}
{\includegraphics[width=8.3cm]{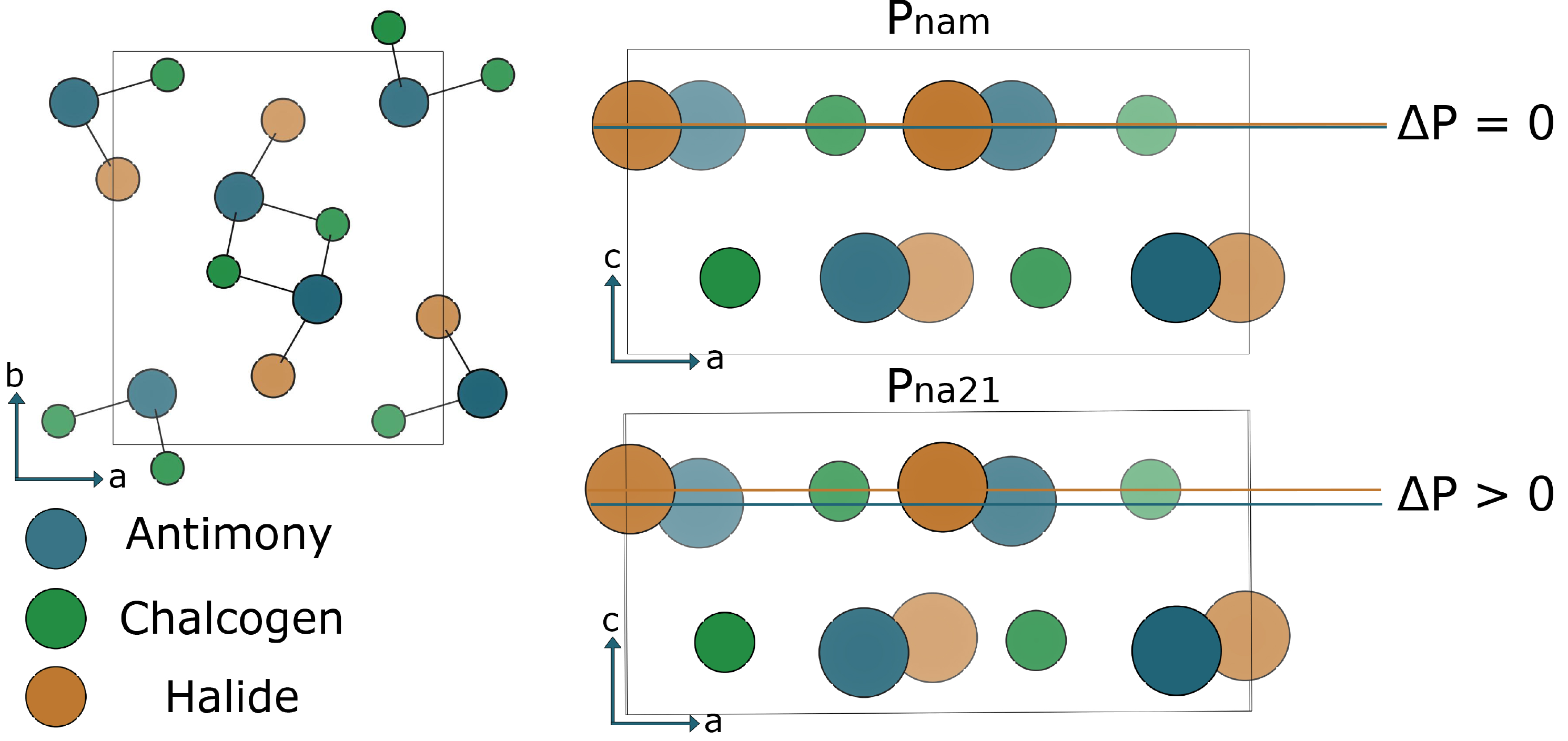}}
\caption{\label{SbSI} Left, view down the (001) direction of the general V-VI-VII structure. Right, schematic of the ferroelectric structural distortion. The centrosymmetric phase $Pnam$ (upper), has no net polarisation. A shift of the Sb sub-lattice results in a breaking of the crystal inversion, and a net macroscopic polarisation $\Delta P$, $Pna21$ (lower).} 
\end{center}
\end{figure}

For much of the early history of photoferroics the archetypal material for the demonstration of photoferroic effects was SbSI\cite{grekov-1969,fridkin-1967,berlincourt-1964,nitsche-1964,fatuzzo-1962}. 
This material has two phases linked by a ferroelectric distortion.
As demonstrated in Figure \ref{SbSI}, a small displacement along the $z$-axis switches between \textit{Pnam} (centrosymmetric, $D_{2h}$) and \textit{Pna21} (non-centrosymmetric, $C_{2v}$) structures. 
The phase behaviour can be linked to the $s^2$ lone pair electrons associated with Sb(III). 
Similar to Pb(II) and Sn(II), the ion can exhibit a second-order Jahn-Teller instability associated with the change from a symmetric to asymmetric coordination environment\cite{walsh-4455}.

The ferroelectric transition, which results in one phase with spontaneous polarisation, means that SbSI is an ideal candidate material for exhibiting both bulk and poly-crystalline photoferroic effects.  
In recent years, this material has been largely overlooked as a potential earth-abundant solar absorber. 
By applying modern electronic structure techniques we asses the utility of SbSI as an absorber layer, and investigate the effects of anion substitution on the electronic properties. 
SbSI exhibits an optical band gap of $\approx2$ eV, the value can be tuned by the choice of chalcogen and halide\cite{alward1978electronic}.

\begin{figure}[t!]
\begin{center}
{\includegraphics[width=8.3cm]{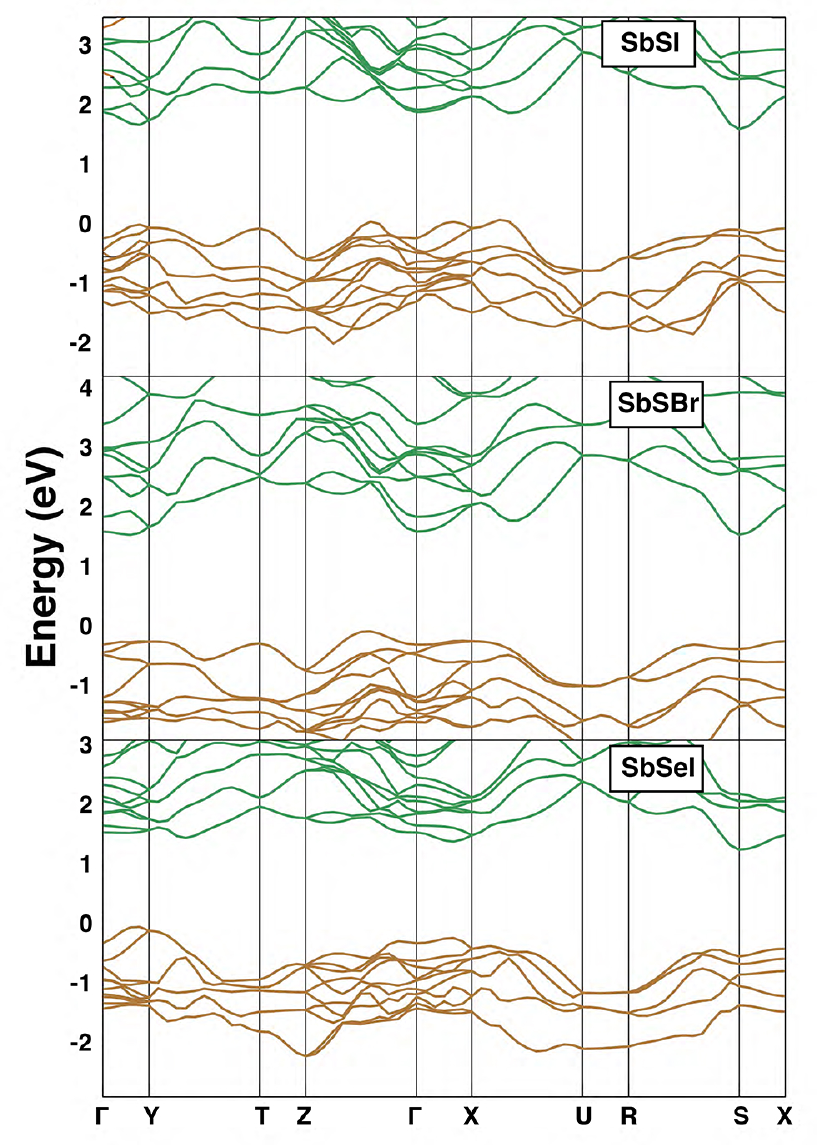}}
\caption{\label{Bands_plots} Energy-momentum band structures of SbSI (upper panel), SbSBr, SbSeI (lower panel); all calculated with GGA-DFT. Valence bands are coloured gold and valence bands are in green. The top of the valence band is set to 0 eV.} 
\end{center}
\end{figure}

For the calculation of the electronic structure properties of SbSI we compare the results of scalar-relativistic generalised-gradient approximation density functional theory (GGA-DFT), hybrid-DFT (HSE06) and hybrid-DFT with spin-orbit coupling (HSE06-SOC). 
The shape of the bands and the  electron and hole effective masses ($m_e^*$ and $m_h^*$) are relatively insensitive to the level of theory: $m_e^*$=0.21, 0.21, 0.22 and $m_h^*$=0.27, 0.27,0.34 for GGA, hybrid and hybrid-SOC, respectively. 
For SbSBr the electron effective mass from GGA (HSE06) is 0.26 (0.23), and for SbSeI the value is 0.52 (0.45). 
The electronic band gap from GGA (1.51 eV) is smaller than for hybrid-DFT (2.14 eV), the further inclusion of spin-orbit coupling reduces the hybrid-DFT value to 1.85 eV. 
For quantitative predictions of the band gaps, relativistic many-body electronic structure theory (e.g. the \textit{QSGW} method) would be required\cite{brivio-2014}.
The electronic and optical band gaps are likely to exhibit a strong temperature dependence owing to the polar nature of the structural phase transition. 

\subsection{Sb(S/Se)X: electronic band structure}
The electronic band structures of SbSI, SbSBr and SbSeI are shown in Figure ~\ref{Bands_plots}. 
All three are indirect bandgap materials and have experimentally reported band gaps of 1.88 eV, 2.20 eV and 1.66 eV, respectively\cite{madelung-04}. 
SbSI has an indirect gap as the top of the valence band (VBM) lies close to the X point and the bottom of the conduction band (CBM) is at the S point. 
It should be noted that the difference between direct and indirect bandgaps is small (0.15 eV); most optical absorption will be direct, and so a thin-film architecture is possible.

Substitution of I by Br results in a change in the electronic band structure. The VBM now occurs between Z and $\Gamma$ and the CBM at $\Gamma$. 
Substitution of S by Se results in another qualitative alteration of the band gap: the VBM is between gamma and Y in the first Brillouin zone, and the CBM is between Y and T. 
The value of the gap is 1.3 eV, close to optimal for solar radiation absorption.

We explain this variation in the electronic structure through the chemical make-up of the bands.
The upper valence band in all cases is composed of chalcogen and halide $p$ orbitals.
The relative contribution from the chalcogen is increased in SbSeI: the lower ionisation potential of Se relative to S explains the band engineering effect, resulting in a gap narrowing. 
Subtle changes in the local environment of Sb are responsible for the change in band extrema, which 
can be associated with the compositional dependence of the stereochemical activity of the Sb 5s$^2$ lone pair electrons.
In contrast, the lower conduction band is comprised of Sb $5p$ orbitals, which are affected by spin-orbit coupling.

The sulpho-halide materials have a small enthalpy difference between the ferroelectric and paraelectric phases ($\Delta$E in Table~\ref{table1}), indicating that transitions between the two phases will be susceptible to the kinds of changes outlined in Section 4.1. 
SbSeI has a larger enthalpy difference, the lower energy ferroelectric phase will be  `locked-in'. 
Berry phase analysis of the polarisation ($\Delta$P in Table~\ref{table1}) indicates that although all three materials have a smaller $\Delta$P than many oxide perovskite structures such as BFO, they all nonetheless posses significant spontaneous electric dipole moments.

\begin{table}[htbp]
  \centering
  \caption{Results of scalar-relativistic GGA-DFT calculations. Band gap (E$_g$ in eV), electron and hole effective masses (m$_e^*$, m$_h^*$), lattice parameters ($a,b,c$ in \AA), polarisation ($\Delta$P in $\mu$C/cm$^2$) and ferroelectric/paraelectric phase energy difference ($\Delta$E in meV per f.u.).}
    \begin{tabular}{lcccccc}
    \hline
          & E$_g$    & m$_e^*$    & m$_h^*$    & $a,b,c$ & $\Delta$P & $\Delta$E \\
    \hline
    SbSI  & 1.51 & 0.21 & 0.27 & 8.5, 10.2, 4.0 & 11 & 59 \\
    SbSBr & 1.57 & 0.26 & 0.57 & 8.2, 9.8, 3.9 & 17 & 2 \\
    SbSeI & 1.29 & 0.52 & 0.24 & 8.3, 11.7, 4.1 & 10 & 376 \\
  \hline
  \hline
    \end{tabular}%
  \label{table1}%
\end{table}

\section{Towards high-efficiency solar energy conversion}

While the main recent focus for solar cells based on ferroelectrics has been on metal oxides, with limited spectral response in the visible range, the consideration of polar chalcogenide and halide semiconductors opens up several new avenues for fundamental research. 

\begin{figure}
\begin{center}
{\includegraphics[width=7.8cm]{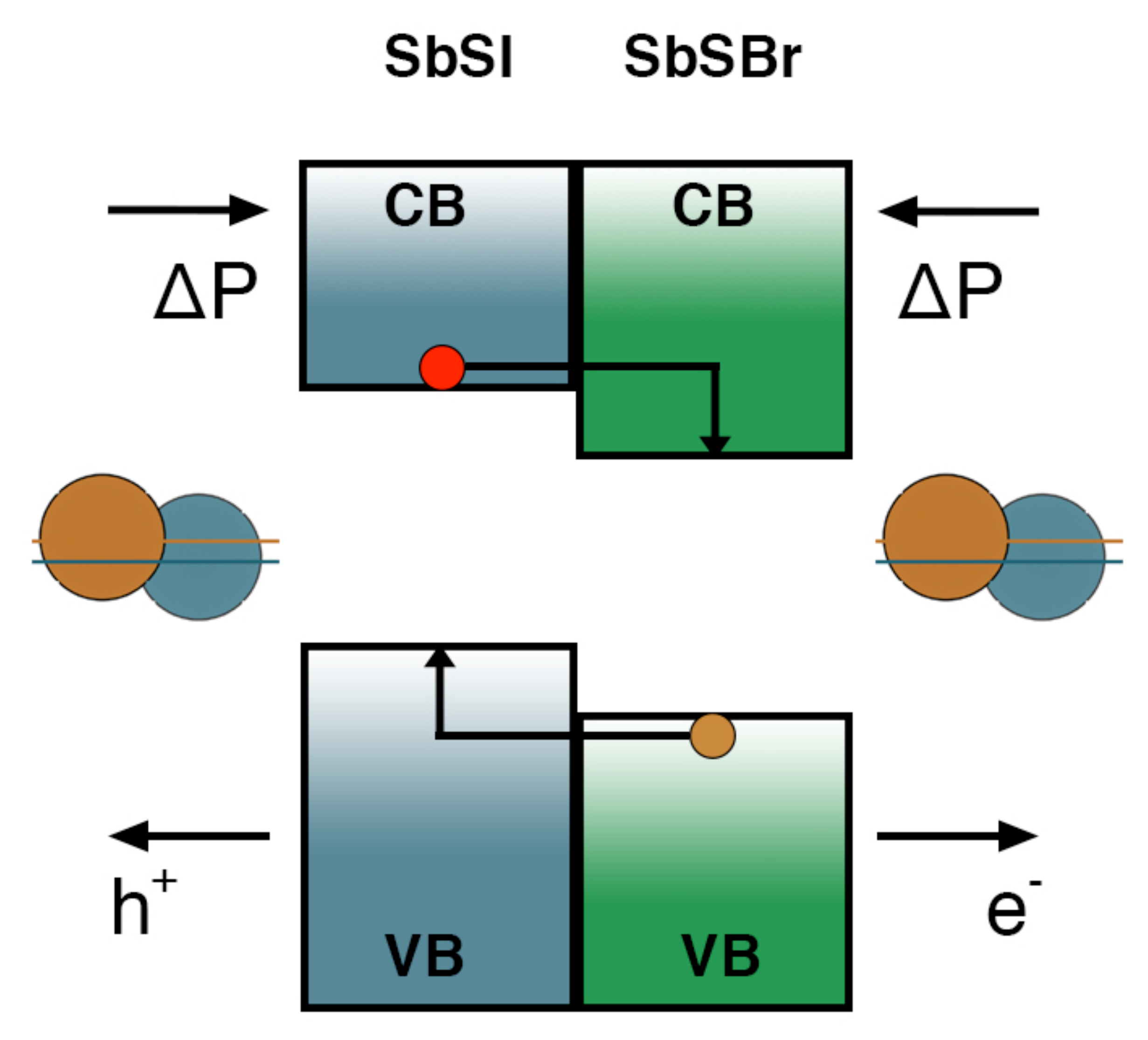}}
\caption{\label{Device_arch} High-efficiency charge separation using a SbSI/SbSBr heterojunction. The band energies of the materials are aligned in a type II offset. The bulk polarisation is in the same direction, so the carriers, which split at the interface, are swept in opposite directions. } 
\end{center}
\end{figure}

Considering the case of the Sb chalcohalides, the effective masses of the charge carriers in all of the materials (Table ~\ref{table1}) suggest high mobility will be possible in good quality crystals with low defect concentrations. 
The closely matched lattice parameters, and the systematic band offsets resulting from chemical substitution, mean that these are ideal candidates for semiconductor heterojunctions. 
For example, epitaxial growth of SbSBr on SbSI, with aligned polarisations. 
In this configuration the combination of type II band offset (driven by the chemistry of the halide ions) and parallel electrical fields could be employed to design efficient charge separation structures; sweeping carrier of opposite charge in opposite directions. Such a device is shown schematically in Figure \ref{Device_arch}.

There is a great opportunity for exploring similar effects in other families of materials. 
For instance, a recent review of metal-organic ferroelectrics highlighted many hybrid metal halide and metal formate compounds that exhibit ferroelectric transitions, with chemical and structural similarities to the hybrid halide perovskite systems\cite{hang2011metal}. 
Mixed anion inorganic compounds, including oxychalcogenides, oxypnictides and chalcopnictides, are of particular interest as the lower symmetry associated with the multi-component systems, coupled with polarisation driven by the electronegativity differences of the constituent anions, ensures that the materials will exhibit complex behaviour. 
A grand challenge is to identify materials with properties similar to the hybrid perovskite s (i.e. light absorption, conductivity, dynamic polarisation, and ease of fabrication), but where Pb is replaced by a more sustainable element.

Another application of ferroelectric materials in PV is for tuning band offsets. The energy offset at junctions between materials
is a major source of efficiency loss in PV devices\cite{Gwinner2011,butler-2012}, ideally there should be Ohmic contact between materials\cite{nelson-2003}. Absolute electron energies are known to be highly sensitive
to external and internal dipoles\cite{walsh-2013p}.
Moreover, the relative positions of valence and conduction bands has previously been shown to depend on surface and interface dipoles\cite{anderson-1990,vdw-1986}.
Simple oxide layers and surface effects have been shown to affect electron energies by as much as 1 eV\cite{Klein2010, butler-2014}, this effect could be even larger from 
a thin film of a polar material, and would allow for the application of alternative, cheaper contacting materials in a range of PV
architectures, for example replacing \ce{In2O3} for organic\cite{butler-2014} and CdTe\cite{ruggeberg-281} devices and replacing silver in silicon PV\cite{butler-2013a}. 

In summary, the bulk properties of ferroelectric materials are important for solar cells, in particular, influencing electron-hole separation and band alignment. 
Beyond these macroscopic effects, photoferroic processes can be significantly enhanced at the nanoscale. 
Understanding and quantifying the interplay between charge carrier distributions, ion transport, and polar structural domains will provide a major challenge for scientists and engineers in this field. 
Furthermore, the interface between polar domains can be as critical as the bulk polarisation itself: the field of domain wall engineering is growing, with many novel and unexpected optoelectronic properties associated with extended defects. 
To paraphrase V. M. Fridkin, \textit{Let us hope that ferroelectric photovoltaics will have a bright future for solar energy generation}.

\section{Acknowledgements}
The work has been funded by EPSRC Grants EP/K016288/1, EP/M009580/1 and EP/J017361/1, with support from the Royal Society and ERC (Grant 277757). 
We acknowledge membership of the UK's HPC Materials Chemistry Consortium, which is funded by EPSRC grant EP/L000202. 

\bibliography{library} 

\end{document}